# Low-temperature synthesis of SmFeAsO$_{0.7}$F$_{0.3-\delta}$ wires with high transport critical current density

Lei Wang, Yanpeng Qi, Dongliang Wang, Zhaoshun Gao, Xianping Zhang, Zhiyu Zhang, Chunlei Wang,  Yanwei Ma[*]

Key Laboratory of Applied Superconductivity, Institute of Electrical Engineering, Chinese Academy of Sciences, P. O. Box 2703, Beijing 100190, China

**Abstract**

Ag-sheathed SmFeAsO$_{0.7}$F$_{0.3-\delta}$ (Sm-1111) superconducting wires were prepared by a one-step solid state reaction at temperatures as low as 850~900 °C, instead of commonly used temperatures of 1150~1250°C. The X-ray diffraction pattern of the as-sintered samples is well indexed on the basis of tetragonal ZrCuSiAs-type structure. We characterized transport critical current density $J_c$ of the SmFeAsO$_{0.7}$F$_{0.3-\delta}$ wires in increasing and subsequently decreasing fields, by a resistive four-probe method. A transport $J_c$ as high as ~1300 A/cm$^2$ at 4.2 K and self field has been observed for the first time in Sm-1111 type polycrystalline superconductors. The $J_c$ also shows a rapid depression in small applied fields as well as a magnetic-history dependence, indicating weak-linked grain boundaries. The low-temperature synthesis method can be very beneficial to fabricating the RE-1111 iron oxynictides in a convenient and safe way.

[*] Author to whom correspondence should be addressed; E-mail: ywma@mail.iee.ac.cn

**Introduction**

The recent discovery of superconductivity in LaFeAsO$_{1-x}$F$_x$ (La-1111) and related compounds, with a highest $T_c$ of ~55 K [1, 2], has triggered worldwide research efforts in both theoretical and experimental aspects for the purpose of understanding the high-$T_c$ mechanism, discovering new superconductors, and investigating magnetic and electronic properties of the new superconductors [3-14]. In particular, the fluorine doped RE-1111 (RE = Sm, Pr and Nd) oxypnictides have drawn great interest, because of their high critical transition temperatures $T_c$ (>50 K), high local critical current density $J_c$ and high upper critical field $H_{c2}$ [15].

Initially, the RE-1111 oxypnictides were synthesized in sealed quartz tubes at temperatures as high as 1150-1250 °C [1, 16], but the synthesis process is non-ideal due to volatility of some components (F and As) and explosion accident. Another method, high pressure synthesis, was done in boron nitride crucibles under high pressure (up to ~5 GPa) and high temperatures (~1200 °C). This method can be used to make fluorine free RE-1111 oxypnictides with a high density, but it is restrict to small sample dimensions [2, 17]. Based on the conventional powder-in-tube method, we have proposed a metal-tube sealed one-step method, which avoid the explosion accident and is versatile for the fabrication of FeAs-based oxypnictides [6]. However, the sintering temperatures are still as high as ~1180 °C [18].

In this paper, we report first synthesis of fluorine doped Sm-1111 wires at temperatures as low as 850~900 °C. Transport critical current density of the SmFeAsO$_{0.7}$F$_{0.3-\delta}$ was characterized in increasing and subsequently decreasing fields, by a resistive four-probe method, and evidence for weak-link behavior were demonstrated.

**Experimental details**

Fluorine doped Sm-1111 polycrystalline were prepared by the one-step method and the details of fabrication process are described elsewhere [19]. Sm filings, Fe powder, Fe$_2$O$_3$ powder, As pieces and SmF$_3$ powder, with a ratio Sm : Fe : As : O : F = 1 : 1 : 1 : 0.7 : 0.3, were well ground in Ar atmosphere, with the aim to achieve a uniform distribution. The final powder was filled in a silver tube (OD: 8 mm, ID: 6.4 mm). The composite were filled in an iron tube (OD: 11.6 mm, ID: 8.2 mm). The filled tube was swaged and drawn down to a wire of 1.95 mm in diameter. Short samples were cut from the as-drawn wires for sintering. The short wires were sintered at 500 °C for 15 hours, heated to 850-900 °C for 40 hours in Ar atmosphere.

Phase identification was characterized by powder X-ray diffraction (XRD) analysis with Cu-K$\alpha$ radiation from 20 to 70°. Resistivity measurements were carried out by the standard four-probe method using a PPMS system. Dc magnetization measurements were performed with a superconducting quantum interference device SQUID magnetometer. The transport critical currents $I_c$ at 4.2 K and its magnetic dependence were evaluated at the High Field Laboratory for Superconducting Materials (HFLSM) in Sendai, Japan, by a standard four-probe resistive method, with a criterion of 1 $\mu$V cm$^{-1}$. The $I_c$ measurement was performed for 2-3 samples to check reproducibility.

**Results and discussion**

X-ray diffraction pattern of the SmFeAsO$_{0.7}$F$_{0.3-\delta}$ wire samples taken after heat-treatment is shown in Fig. 1. As is seen from the figure, most of peaks as noted can be indexed on the basis of tetragonal ZrCuSiAs-type structure, with a small amount of impurity phases, ensuring the Sm-1111 phase can be synthesized at 850-900 °C, about 300 °C lower than those (1150~1200 °C) reported previously [16-18]. In the present study, we adopted 0.3-$\delta$ for the fluorine concentration because fluorine evaporated during the heat-treatment process. The lattice parameters were derived from the X-ray diffraction pattern using a X'Pert Plus program, and the resultant values are $a$ = 0.3933 nm and $c$ = 0.8477 nm, respectively. Compared to the undoped SmFeAsO ($a$ = 0.3940 nm, $c$ = 0.8450 nm), both a- and c- axis were reduced, indicating fluorine was successfully adopted into Sm-1111 lattice [20].

Figure 2 shows normalized resistivity versus temperature measurement of the SmFeAsO$_{0.7}$F$_{0.3-\delta}$ specimens. In agreement with previous reports, SmFeAsO$_{0.7}$F$_{0.3-\delta}$ is characterized by a nearly linear decrease of resistivity with temperature. The onset of superconductivity occurs at 41 K, and zero resistivity is achieved at 31 K (upper inset). The $T_c$ is consistent with the value reported by Chen's group, for which the sintering temperatures are 1160~1200 °C [16]. In addition, a slight drop of resistivity was observed above 48 K. Since the critical transition temperature $T_c$ of SmFeAsO$_{1-x}$F$_x$ is strongly dependent on fluorine concentration, we suggest that Sm-1111 phase with higher $T_c$ can be synthesized at such low temperatures by depressing fluorine volatility.

The lower inset of Fig. 2 displays temperature dependence of zero-field-cooled (ZFC) and field-cooled magnetization for the SmFeAsO$_{0.7}$F$_{0.3-\delta}$ specimen measured under 20 Oe. The diamagnetic signal on zero-field-cooling appears near 40 K, and significantly increases below 30 K, which is well consistent with the resistance result. The FC and ZFC measurements yield maximum signals of ~2% and ~80% of complete flux expulsion, respectively. The small diamagnetic signal on field-cooling indicates that the specimen shows strong pinning of trapped vortices.

Global critical current densities, $J_c$'s, of the polycrystalline FeAs-based superconductors, have been investigated by many groups, through remanent magnetization, magneto-optical imaging, Campbell's method and resistive four-probe method, and evidence for granular behavior were presented [7-12]. The global $J_c$ have generally been restricted to values <10$^4$ A/cm$^2$ at 4.2 K in self field, and a rapid depression was observed by applying small magnetic fields (~0.1 T) [10]. Recently, an experimental result showed that weak-link grain boundaries are characteristic of iron pnictides, and they ascribed this to competing orders, low carrier density and unconventional pairing symmetry [21]. Another study argued that the origin of low $J_{cb}$ is the tensile strain generated by dislocations located along the boundary [22]. Further study on the features of transport critical current density in polycrystalline oxypnictides, especially by a standard four-probe method, could be much helpful for understanding the weak-link mechanism in iron-based superconductors.

We evaluated transport critical currents, $I_c$'s, of the SmFeAsO$_{0.7}$F$_{0.3-\delta}$ wires by a standard four-probe method in applied fields up to 14 T. Two typical original current-voltage plots were shown in the inset of Fig. 3, and the zero resistive currents of ~13 A and ~1 A, were clearly seen for self field and 14 T, respectively. A repeated measurement was carried out right after a virgin measurement at 0 T, and the resultant plot was almost the same as the virgin one, suggesting that the transport property of the SmFeAsO$_{0.7}$F$_{0.3-\delta}$ superconducting wires are not affect by current history.

Transport critical current density $J_c$ as a function of field for Sm-1111 wires is presented in Fig. 3. A transport $J_c$ as high as ~1300 A/cm$^2$ at 4.2 K and self field has been observed for the first time in 1111-type polycrystalline superconductors. The $J_c$ values show a strong field dependence in low field region (e.g. $J_c$ (0 T) ≥ 10 $J_c$ (0.5 T)), and this is in analogue to that of sintered YBCO, which shows an intrinsic weak-link behavior. In high field region, the $J_c$ is almost field independent, constant at ~80 A/cm$^2$ between 0.5 T and 14 T. In Sr$_{0.6}$K$_{0.4}$Fe$_2$As$_2$, a similar field dependence of $J_c$ was observed, but the $J_c$ value is much lower [23]. The transport $J_c$ in Sm- and Nd-1111 bulk superconductors, which were fabricated under 6 GPa at 1250 $^o$C, have been evaluated previously, and it was reported that the defects at grain boundaries, such as impurity phases and cracks, produce a multiply connected current-blocking network [10]. Clearly, the impurity phases, peaks of which were observed in the XRD pattern, were thought to be harmful to the transport capability, and a higher transport $J_c$ should be expected for SmFeAsO$_{0.7}$F$_{0.3-\delta}$ superconductors with improved raw powder and sintering process.

After increasing the field monotonically to 14 T, transport $J_c$ was also measured as a function of decreasing field until 0 T. An important point to note is the increased value of $J_c$ in the region above 0.4 T, compared with the data for the virgin curve. The $J_c$ is seen to peak on decreasing the field at 0.4 T, so that for 0 T, the $J_c$ value is substantially reduced as compared to the virgin measurement. The hysteretic effects are supposed to be related to weak-linked grain boundaries as well as penetration of flux into strong pinning intra-granular regions, and that the presence of intra-granular critical currents enhances inter-granular critical currents when the applied field is reduced from higher values. The peak in the transport $J_c$ for decreasing fields, is generally attributed to compensation of the applied field and by the effect of paramagnetic intra-granular currents generated in decreasing field [24].

Results confirm that synthesis of fluorine doped Sm-1111 iron oxypnictide can be achieved with a low temperature one-step method. Because of its high similarity with other RE-1111 oxypnictides, the low temperature synthesis can be extended to many different materials, such as Nd-, Pr-, Ce- and La- 1111. However, the main advance is that it is not necessary to afford high temperatures of ~1200 $^{o}$C for synthesis iron oxypnictides. This becomes more important when one considers the high As vapour-pressure and the resultant explosion accident. Although the transport critical current density $J_c$ in polycrystalline iron oxypnictides is limited by the weak links between grains, higher $J_c$ values could be expected in pure and textured materials, which is the on-going research. This low-temperature synthesis would give further encouragement to the development of the RE-1111 FeAs-based superconductors, with a highest $T_c$ of 55 K, for potential applications.

**Conclusions**

We demonstrated a Sm-1111 synthesis temperature as low as ~900 $^{o}$C. Such a temperature is far lower than those reported previously, and is comparable with the synthesis temperatures of 122 iron pnictides (850-900 $^{o}$C). The low temperature synthesis method reported here can be applied to other RE-1111 oxypnictides, such as Nd-, Pr-, Ce- and La- 1111, to synthesize a variety of elemental polycrystalline superconductors and single crystals, using appropriate precursors and heat-treatment program.

**Acknowledgements**

The authors thank Profs. Satoshi Awaji, Haihu Wen, Liye Xiao and Liangzhen Lin for their help and useful discussions. This work was partially supported by the Beijing Municipal Science and Technology Commission under Grant No. Z09010300820907, National Science Foundation of China (grant no. 50802093) and the National '973' Program (grant no. 2006CB601004).

**Captions**

Figure 1 X-ray diffraction pattern of the polycrystalline SmFeAsO$_{0.7}$F$_{0.3-\delta}$ wires taken after heat-treatment.

Figure 2 Temperature dependence of normalized resistivity for the polycrystalline SmFeAsO$_{0.7}$F$_{0.3-\delta}$ wires; Upper inset: Details near $T_c$; Lower inset: Temperature dependence of DC susceptibility for the polycrystalline SmFeAsO$_{0.7}$F$_{0.3-\delta}$.

Figure 3 Transport critical current density $J_c$ as a function of increasing as well as decreasing fields for SmFeAsO$_{0.7}$F$_{0.3-\delta}$ superconducting wires. Inset: Original current-voltage plots at self field and 14 T.

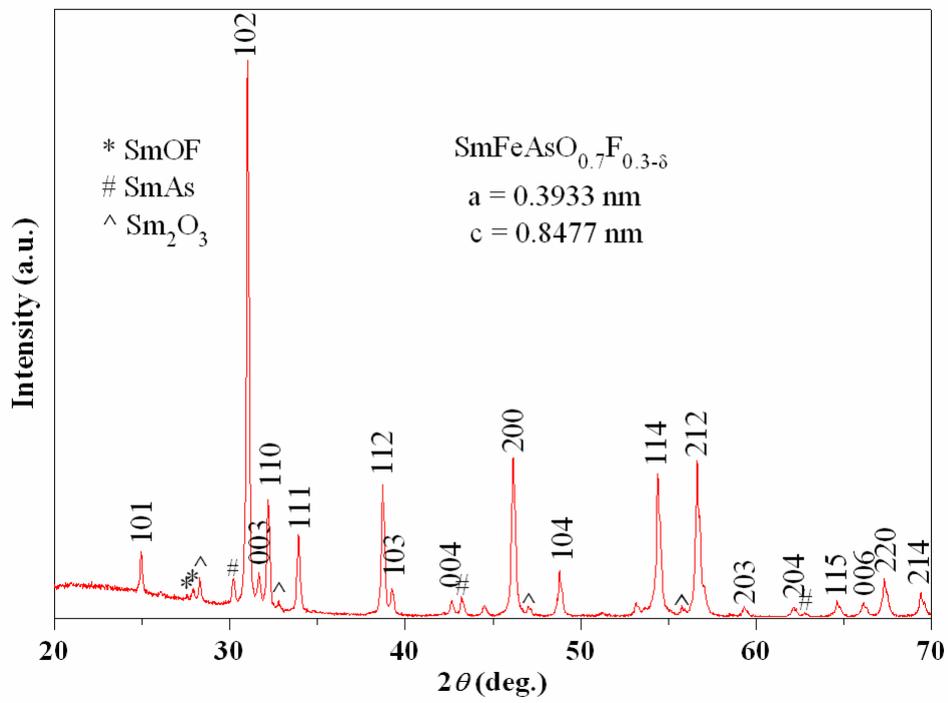

Figure 1 Wang et al.

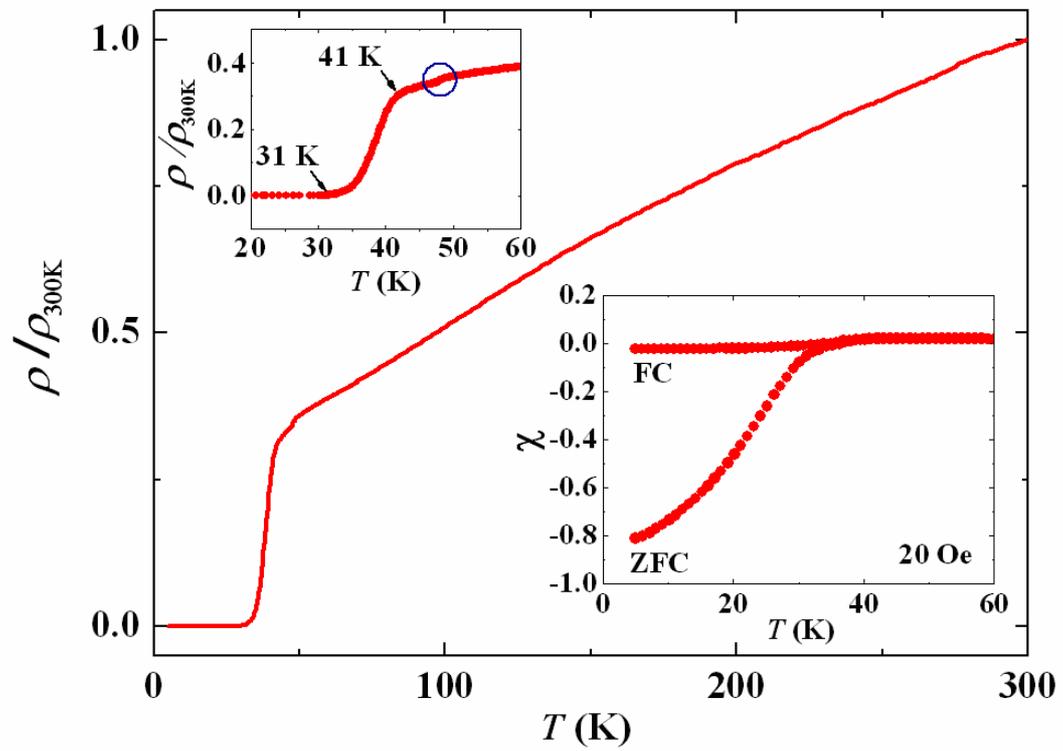

Figure 2 Wang et al.

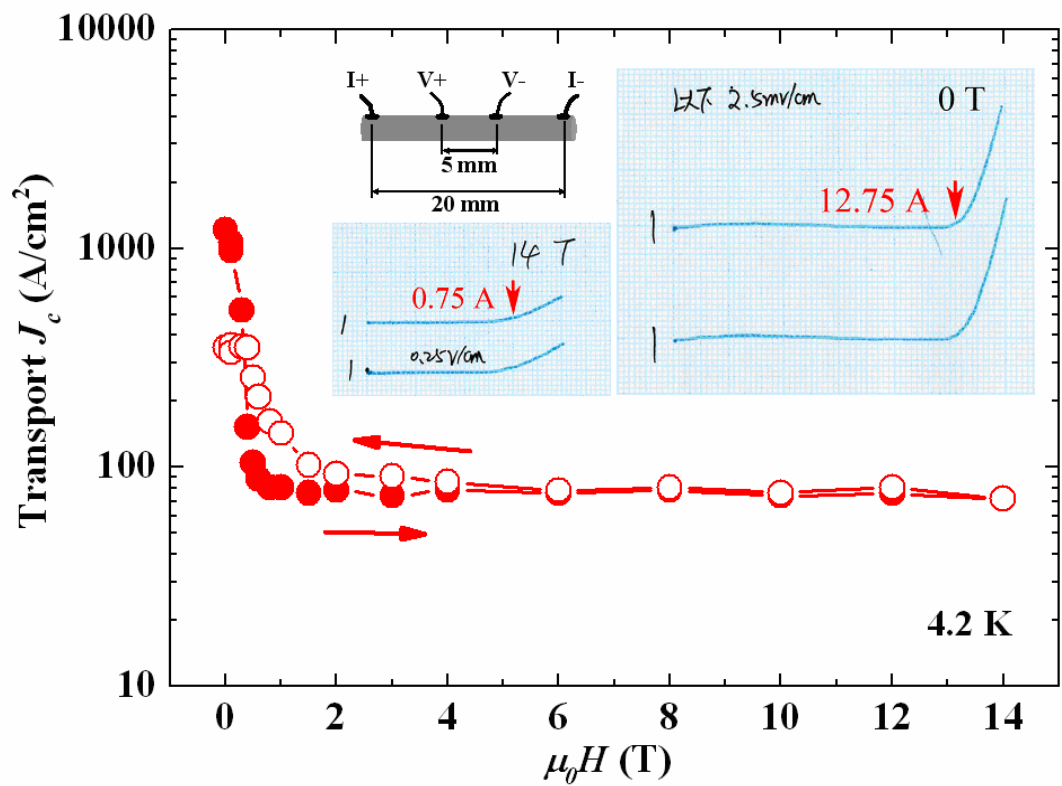

Figure 3 Wang et al.